\documentclass[a4j,12pt]{article}
\usepackage{amsmath,amsthm,amssymb}
\usepackage{latexsym}
\usepackage{amsmath}
\usepackage{amssymb}
\usepackage[pdftex]{graphicx}
\usepackage{wrapfig}
\usepackage{cases}
\usepackage{fancybox}
\usepackage{bm}
\usepackage{multirow}

\newcommand{\Gcenter}[2]{
	\dimen0=\ht\strutbox%
	\advance\dimen0\dp\strutbox%
	\multiply\dimen0 by#1%
	\divide\dimen0 by2%
	\advance\dimen0 by-.5\normalbaselineskip
	\raisebox{-\dimen0}[0pt][0pt]{#2}}%

\begin{document}
\baselineskip=1.2\baselineskip

\pagestyle{plain}
\begin{center}
{\large \bf Visualizing departures from marginal homogeneity for square contingency tables with ordered categories}

\vspace{1cm}
Satoru Shinoda${}^{1}$, Takuya Yoshimoto${}^{2}$ and  Kouji Tahata${}^{3}$\\
\vspace{1cm}
${}^{1}${\it Department of Biostatistics, Yokohama City University, School of Medicine, Japan}\\
${}^{2}${\it Biometrics Department, Chugai Pharmaceutical Co., Ltd., Japan}\\
${}^{3}${\it Department of Information Sciences, Faculty of Science and Technology, Tokyo University of Science, Japan}\\
\vspace{0.5cm}
E-mail: shinoda.sat.cg@yokohama-cu.ac.jp\\
\vspace{0.5cm}
{\bf Abstract}
\end{center}

Square contingency tables are a special case commonly used in various fields to analyze categorical data. 
Although several analysis methods have been developed to examine marginal homogeneity (MH) in these tables, existing measures are single-summary ones. 
To date, a visualization approach has yet to be proposed to intuitively depict the results of MH analysis. 
Current measures used to assess the degree of departure from MH are based on entropy such as the Kullback-Leibler divergence and do not satisfy distance postulates. 
Hence, the current measures are not conducive to visualization.
Herein we present a measure utilizing the Matusita distance and introduce a visualization technique that employs sub-measures of categorical data. 
Through multiple examples, we demonstrate the meaningfulness of our visualization approach and validate its usefulness to provide insightful interpretations.

\vspace{0.5cm}
\begin{flushleft}
\\
\end{flushleft}

\newpage
\noindent \textbf{\large 1. Introduction}

Numerous research areas employ categorical data analysis. 
Such data is summarized in a contingency table (see e.g., Agresti, 2013; Kateri, 2014).
A special case is a square contingency table where the row and column variables have the same ordinal categories. 
When we cannot obtain data as continuous variables for the evaluation of the efficacy and safety/toxicity of treatments in clinical studies, ordered categorical scales are used alternatively.

For example, Sugano $et~al.$ (2012) conducted a clinical study where they examined the modified LANZA score (MLS) after 24 weeks’ treatment with esomeprazole 20 mg once daily or a placebo.
The MLS is a popular evaluation scale with five stages (from 0 to +4) and is used for clinical evaluations of gastroduodenal mucosal lesions.
Table 1 shows a square contingency table that summarizes the location shift of the MLS from pre-treatment to post-treatment for each patient.
Such research is interested in whether the treatment effect tends to improve or worsen after an intervention relative to before the intervention.
Thus, the evaluation is interested in the similarity from marginal homogeneity (MH), but not independence.
Stuart (1955) introduced the MH model to indicate homogeneity with respect to two marginal distributions.
We are also interested in the structure of inhomogeneity of the two marginal distributions when the MH model does not hold.
This is because we are more interested in the deviation between the pre-treatment and post-treatment marginal distributions (i.e., intervention results) than whether the MH model that represents the structure shows an equal marginal distribution for the data in Table 1.
Consequently, our strategy is to estimate measures representing the degree of departure from MH.
Measures must quantify the differences in probability distributions, mainly using information divergences such as Kullback-Leibler divergence or power-divergence.

\begin{table}[h]
\small\sf\centering
\caption{Shift analysis data of MLS after treatment for 24 weeks with esomeprazole 20 mg once daily or a placebo.\label{T1}}
\begin{tabular}[!bh]{cccccc} \hline
& \multicolumn{5}{c}{Baseline} \\  \cline{2-6} 
Study end	& 0	& +1	& +2	& +3	& +4	\\ \hline
\multicolumn{5}{l}{(a) Esomeprazole 20~mg once daily}\\ 
0		& 78	& 9	& 26	& 3	& 1	\\
+1		& 1	& 5	& 6	& 4	& 0	\\
+2		& 9	& 1	& 10	& 3	& 1	\\
+3		& 1	& 0	& 1	& 0	& 0	\\
+4		& 3	& 0	& 1	& 1	& 2	\\ \hline
\multicolumn{5}{l}{(b) Placebo}\\
0		& 41	& 2	& 19	& 0	& 0	\\
+1		& 8	& 0	& 4	& 0	& 0	\\
+2		& 12	& 4	& 14	& 3	& 0	\\
+3		& 0	& 1	& 1	& 3	& 0	\\
+4		& 29	& 7	& 11	& 6	& 0	\\ \hline
\end{tabular}
\end{table}

To this end, Tomizawa, Miyamoto and Ashihara (2003) proposed a measure using the marginal cumulative probability for square contingency tables with ordered categories.
This measure ranges from 0 to 1 and directly represents the degree of departure from MH.
However, it cannot distinguish the direction of degree of departure.
The two marginal distributions are interpreted as equal (no intervention effect) when the value is 0.
When the values are greater than 0, an improvement is indistinguishable from a worsening effect.
Yamamoto, Ando and Tomizawa (2011) proposed a measure, which lies between -1 and 1, to distinguish the directionality. 
This measure cannot represent the degree of departure directly from MH. 
Even if the value of the measure is 0, the marginal distribution cannot be exactly interpreted as having no intervention effect. 
To simultaneously analyze the degree and directionality of departure from MH, Ando, Noguchi, Ishii and Tomizawa (2021) proposed a two-dimensional visualized measure that combines the measure proposed by Tomizawa $et~al.$ (2003) and the measure proposed by Yamamoto $et~al.$ (2011). 
They also considered visually comparing the degrees of departure from MH in several tables because their measure is independent of the dimensions (i.e., number of categorical values) and sample size. 
Appendix 1 explains the main points of the above measures.

These measures proposed by Tomizawa $et~al.$ (2003), Yamamoto $et~al.$ (2011) and Ando $et~al.$ (2021) are single-summaries.
They are expressed using the sub-measure weights at each categorical level. 
For a given category level, different behaviors cannot be distinguished as a single-summary measure. 
The artificial data examples in the data analysis section provide specific situations. 
Hence, a single-summary-measure may overlook different behaviors in a given categorical level.
To address this limitation, we apply visualization as a method utilizing sub-measures defined at each category level. 
This visualization also assumes that satisfying distance postulates can achieve a natural interpretation. 
To date, a measure for ordered categories does not exist because the Kullback-Leibler divergence or power-divergence used in existing measures do not satisfy the distance postulates. 
Therefore, we consider a measure using the Matusita distance to capture the discrepancy between two probability distributions while satisfying the distance postulates (see Matusita, 1954, 1955; Read and Cressie, 1988, p.112).

Both academia and general society employ methods to visualize quantitative data. 
Examples include pie charts, histograms, and scatterplots. 
Although visualizing categorical data has attracted attention recently, different visualization techniques from those for quantitative data are necessary (see, e.g., Blasius and Greenacre, 1998; Friendly and Meyer, 2015; Kateri, 2014).
Visualization of categorical data has two main objectives: revealing the characteristics of the data and intuitively understanding analysis results (Friendly and Meyer, 2015). 
Methods for the former include the ``mosaic plot'' and ``sieve diagram'' (see e.g., Friendly, 1995; Hartigan and Kleiner, 1981, 1984; Riedwyl and Sch\"{u}pbach, 1983, 1994).
Methods for the latter include the ``fourfold display'' for odds ratios and the  ``observer agreement chart'' for Cohen’s $\kappa$ (see e.g., Bangdiwala 1985, 1987; Fienberg, 1975; Friendly, 1994). 
Although the visualization objectives for categorical data may vary, they share common techniques: (i) separating data by categorical levels and (ii) adjusting the size of figure objects based on the frequency of each cell.

Our research aims to realize a visualization for an intuitive understanding of the analysis results for MH. 
To date, such a visualization has yet to be proposed. 
Although the ``mosaic plot'' and ``sieve diagram'' can be applied to square contingency tables, they are not suitable for examining the structure of MH. 
These visualizations are designed to observe the data itself and identify features or patterns without making hypotheses before analyzing the data. 
Therefore, our proposed visualization provides an intuitive understanding of the structure of MH using categorical data visualization techniques (i) and (ii).

This paper conducts a comprehensive analysis of the degree and directionality of departure from MH for square contingency tables with ordered categories. 
Our approach has two components: (i) measures to quantify the degree of departure of MH using information divergence satisfying distance postulates and (ii) a visualization technique designed for categorical data.

The rest of this paper is organized as follows. 
Section 2 defines the proposed measure and visualization. 
Section 3 derives an approximated confidence interval for the proposed measure. 
Section 4 provides examples of the utility for the proposed measure and visualization. 
Section 5 presents the discussion. 
Finally, Section 6 closes with concluding remarks.
\\

\noindent \textbf{\large 2. Proposed measure and visualization}

Here, we detail the proposed measure and visualization. 
Section 2.1 explains the probability structure of the MH model using formulas. 
Section 2.2 defines the sub-measures and single-summary-measure expressed using weights for the sub-measures at each categorical level along with the properties of the proposed measure. 
Section 2.3 details the visualization of the proposed measures.
\\

\noindent \textbf{\large 2.1. MH model}

Consider an $r \times r$ square contingency table with the same row and column ordinal classifications.
Let $X$ and $Y$ denote the row and column variables, respectively, and let Pr$(X = i , Y = j) = p_{ij}$ for $i = 1, \ldots , r; j = 1, \ldots , r$.

The MH model can be expressed with various formulas.
For example, the MH model is expressed as
\[
p_{i \cdot} = p_{\cdot i} \quad {\rm for} ~ i = 1, \ldots , r,
\]
where $p_{i \cdot} = \sum^r_{t=1}p_{it}$ and $p_{\cdot i} = \sum^r_{s=1}p_{si}$. 
See e.g., Stuart (1955) and Bishop, Fienberg and Holland (1975, p.294).
This indicates that the row marginal distribution is identical to the column marginal distribution.

To consider ordered categories, the MH model can be expressed using the marginal cumulative probability as
\[
F_{1(i)} = F_{2(i)}  \quad {\rm for} ~ i = 1, \ldots , r-1,
\]
where $F_{1(i)} = \sum^i_{s=1} p_{s \cdot} = {\rm Pr}(X \leq i)$ and $F_{2(i)} = \sum^i_{t=1} p_{\cdot t} = {\rm Pr}(Y \leq i)$.
The MH model can also be expressed as
\[
G_{1(i)} = G_{2(i)} \quad {\rm for} ~ i = 1, \ldots , r-1,
\]
where $G_{1(i)} = \sum^i_{s=1} \sum^r_{t=i+1} p_{st} = {\rm Pr}(X \leq i, Y \geq i+1)$ and $G_{2(i)} = \sum^r_{s=i+1} \sum^i_{t=1} p_{st} = {\rm Pr}(X \geq i+1, Y \leq i)$.
Furthermore, the MH model can be expressed as
\[
G^c_{1(i)} = G^c_{2(i)} \left( = \frac{1}{2} \right) \quad {\rm for} ~ i = 1, \ldots , r-1,
\]
where
\[
G^c_{1(i)} = \frac{G_{1(i)}}{G_{1(i)} + G_{2(i)}},
\quad
G^c_{2(i)} = \frac{G_{2(i)}}{G_{1(i)} + G_{2(i)}}.
\]
The MH model states that the conditional probability of $X \leq i$ is given if either $X$ or $Y \leq i$ and the other $\geq i+1$ is equal to the conditional probability that $Y \leq i$ for the same conditions.
\\

\noindent \textbf{\large 2.2. Measure of departure from MH}

Several measures have been proposed for various formulas of the MH model. 
Here, we consider a measure that is independent of the diagonal probabilities because the MH model does not have constraints on the main-diagonal cell probabilities. 
For instance, Tomizawa $et~al.$ (2003) and Yamamoto $et~al.$ (2011) proposed measures that do not depend on the diagonal probabilities.

First, we consider a sub-measure satisfying the distance postulates. 
Assuming that $G_{1(i)} + G_{2(i)} \neq 0$, the degree of departure from MH at each categorical level $i$ ($i=1, \ldots, r-1$) is given as
\[
\gamma_i = \left[ \frac{2+\sqrt{2}}{2} \left( \upsilon_{1(i)}^2 + \upsilon_{2(i)}^2 \right)  \right]^{\frac{1}{2}},
\]
where
\[
\upsilon_{1(i)} = \sqrt{G^c_{1(i)}} - \sqrt{\frac{1}{2}},
\quad
\upsilon_{2(i)} = \sqrt{G^c_{2(i)}} - \sqrt{\frac{1}{2}}.
\]
The sub-measure $\gamma_i$ has the following characteristics:
\begin{enumerate}
\item[(i)] $0 \leq \gamma_i \leq 1$
\item[(ii)] $\gamma_i = 0$ if and only if $G^c_{1(i)} = G^c_{2(i)} (= 1/2)$
\item[(iii)] $\gamma_i = 1$ if and only if $G^c_{1(i)} =1$ (then $G^c_{2(i)} = 0$) or $G^c_{1(i)} =0$ (then $G^c_{2(i)} = 1$)
\end{enumerate}
The sub-measure $\gamma_i$ is the Matusita distance between $\left( G^c_{1(i)}, G^c_{2(i)} \right)$ and $\left( \frac{1}{2}, \frac{1}{2} \right)$, and satisfies all three distance postulates. 
When the value of the sub-measure is 0, it means the marginal cumulative probabilities are equivalent until categorical level $i$.
The value of the sub-measure increases as the separation between the marginal cumulative distributions increases. 
The separation is maximized when the value of the sub-measure is 1. 
Noting that a distance $d$ is defined on a set $W$ if for any two elements $x, y \in  W$, a real number $d(x, y)$ is assigned that satisfies the following postulates:
\begin{enumerate}
\item[(i)] $d(x, y) \geq  0$ with equality if and only if $x=y$;
\item[(ii)] $d(y, x) = d(x, y)$;
\item[(iii)] $d(x, z) \leq d(x, y) + d(y, z)$ for $x, y, z \in  W$ (the triangle inequality).
\end{enumerate}
See also Read and Cressie (1988, p.111).
Then the power-divergence $I^{(\lambda)}$ (especially, the Kullback-Leibler divergence $I^{(0)}$) does not satisfy postulates (ii) and (iii). 
The Matusita distance, which is the square root of $I^{(-\frac{1}{2})}$, satisfies all three postulates. 

Assuming that $\{ G_{1(i)} + G_{2(i)} \neq 0 \}$, we consider a measure using sub-measure $\gamma_i$ to represent the degree of departure from MH, which is given as
\[
\Gamma = \sum^{r-1}_{i=1} \left( G^{\ast}_{1(i)} + G^{\ast}_{2(i)} \right) \gamma_i,
\]
where
\[
\Delta = \sum^{R-1}_{i=1} \left( G_{1(i)} + G_{2(i)} \right),
\]
and
\[
G^{\ast}_{1(i)} = \frac{G_{1(i)}}{\Delta},
\quad
G^{\ast}_{2(i)} = \frac{G_{2(i)}}{\Delta},
\]
for $i=1, \ldots, r-1$.
The measure $\Gamma$ has the following characteristics:
\begin{enumerate}
\item[(i)] $0 \leq \Gamma \leq 1$
\item[(ii)] $\Gamma = 0$ if and only if the MH model holds
\item[(iii)] $\Gamma = 1$ if and only if the degree of departure from MH is a maximum, in the sense that  $G^c_{1(i)}=1$ (then $G^c_{2(i)}=0$) or $G^c_{1(i)}=0$ (then $G^c_{2(i)}=1$), for $i = 1, \ldots, r-1$
\end{enumerate}
Thus, this measure is the weighted sum of the Matusita distance for the two distributions $\left( G^c_{1(i)}, G^c_{2(i)} \right)$ and $\left( \frac{1}{2}, \frac{1}{2} \right)$.
\\

\noindent \textbf{\large 2.3. Visualization of the proposed measure}

To visualize the proposed measure, we used the techniques for visualizing categorical data. 
First, for the fixed $i$ ($i=1, \ldots, r-1$), $\gamma_i$, which represents the relationship between $G^c_{1(i)}$ and $G^c_{2(i)}$, is defined by the following steps:
\begin{enumerate}
\item[(i)] Plot the $x$-axis is $G^c_{1(i)}$ and the $y$-axis is $G^c_{2(i)}$ point for each $\left( G^c_{1(i)}, G^c_{2(i)} \right)$ coordinate
\item[(ii)] Adjust the point size according to the weight $\left( G^{\ast}_{1(i)} + G^{\ast}_{2(i)} \right)$ 
\item[(iii)] Display the value of $\gamma_i$ as a text label at each $\left( G^c_{1(i)}, G^c_{2(i)} \right)$ point
\item[(iv)] Color the points red when $\left(G^c_{1(i)} < G^c_{2(i)}\right)$ and blue when $\left( G^c_{1(i)} \geq G^c_{2(i)}\right)$
\item[(v)] Draw the dashed line within the diagonal point’s range of movement and color the dashed line using the same rules
\end{enumerate}
Therefore, the top-left side is red, while the bottom-right side is blue with respect to the point $\left( \frac{1}{2}, \frac{1}{2} \right)$ in the visualization. 
Table 2 shows a visualization image.
\begin{table}[h]
\small\sf\centering
\caption{True cell probabilities in a $6 \times 6$ square contingency table.\label{T2}}
\begin{tabular}[!bh]{ccccccc} \hline
	& 1	 	& 2 		& 3		& 4		& 5		& 6		\\ \hline
1	& 0.000	& 0.031	& 0.219	& 0.031	& 0.031	& 0.000	\\
2	& 0.000	& 0.000 	& 0.031	& 0.031	& 0.031	& 0.000	\\
3	& 0.000	& 0.031	& 0.000	& 0.031	& 0.031	& 0.000	\\
4	& 0.000	& 0.031	& 0.031	& 0.000	& 0.031	& 0.000	\\
5	& 0.000	& 0.031	& 0.031	& 0.031	& 0.000	& 0.000	\\
6	& 0.000	& 0.031	& 0.031	& 0.219	& 0.031	& 0.000	\\ \hline
\end{tabular}
\end{table}
Table 3 presents the necessary information to visualize Table 2, including $G^c_{1(i)}$ and $G^c_{2(i)}$ used for the coordinates of the point, the weight used for the point size, and the sub-measure $\gamma_i$ used for the text label.
\begin{table}[h]
\small\sf\centering
\caption{Visualization values of $\gamma_i$.\label{T3}}
\begin{tabular}[!bh]{ccccc} \hline
$i$	& $x$-axis	& $y$-axis	& size		& label	\\ \hline
1	& 1.000	& 0.000	& 0.156	& 1.000	\\
2	& 0.750	& 0.250	& 0.250	& 0.341	\\
3	& 0.500	& 0.500	& 0.188	& 0.000	\\
4	& 0.250	& 0.750	& 0.250	& 0.341	\\
5	& 0.000	& 1.000	& 0.156	& 1.000	\\ \hline
\end{tabular}
\end{table}
Step 1 visualizes each level $i$. 
As an example, Figure 1 depicts how $\gamma_i$ is visualized at level $i=1$.
\begin{figure}[h]
\centering
\includegraphics{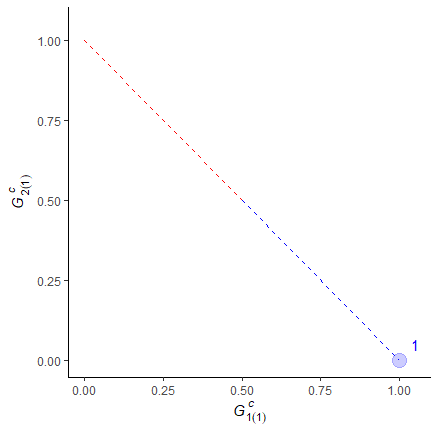}
\caption{Visualization of $\gamma_1$.}
\end{figure}

Next, we provide additional definitions to integrate each $\gamma_i$ in step 1 into one figure:
\begin{enumerate}
\item[(i)] Consider the $x$-axis as $i$ for $G^c_{1(i)}$ and the $y$-axis as $i$ for $G^c_{2(i)}$
\item[(ii)] Place the figure of $\gamma_i$ on the diagonal
\end{enumerate}
Figure 2 shows the integrated figure using the example from Table 2 in step 2 according to the definition of the proposed visualization.
\begin{figure}[h]
\centering
\includegraphics{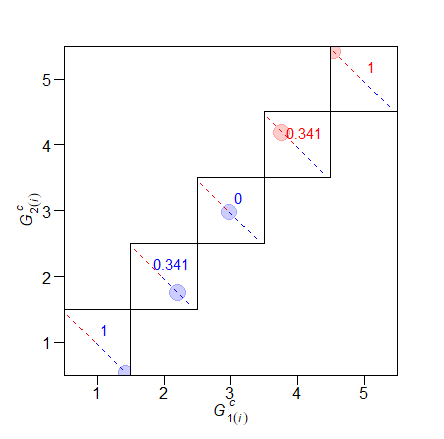}
\caption{Proposed visualization to provide an intuitive understanding of the structure of MH}
\end{figure}

The visualization of the proposed measure using the categorical data methods has the following benefits. 
First, the visualization provides information about each $i$, allowing trends in MH to be identified in a square contingency table. 
Since the figure visualizes each $\gamma_i$, points do not overlap even if their coordinates are close. 
Thus, points are easily identifiable. 
It is important to visualize each $\gamma_i$ separately since each one is assumed to be nearly the same value.

Ando $et~al.$ (2021) used a Kullback-Leibler divergence-type measure, but the Kullback-Leibler divergence does not satisfy the distance postulates. 
To naturally interpret the point distances in the figure, the distance postulates must be satisfied. (Section 4.1.1. gives a specific example). 
Additionally, the proposed visualization can be considered as utilizing sub-measures.
\\

\noindent \textbf{\large 3. Approximate the confidence interval for the measure}

Let $n_{ij}$ denote the observed frequency in the $i$th row and $j$th column of a table ($i =1, \ldots, r;~j = 1, \ldots, r$).
The sample version of $\Gamma$ (i.e., $\hat{\Gamma}$) is given by $\Gamma$ in which \{$p_{ij}$\} is replaced by \{$\hat{p}_{ij}$\}, where $\hat{p}_{ij} = n_{ij}/n$ and $n = \sum \sum n_{ij}$.
It should be noted that the sample version of $G^c_{k(i)}$, $\gamma_i$ and $F_{k(i)}$, which are $\hat{G}^c_{k(i)}$, $\hat{\gamma}_i$ and $\hat{F}_{k(i)}$, respectively, are given in a similar manner ($i=1, \ldots, r-1;~k=1, 2$).
Given that \{$n_{ij}$\} arises from a full multinomial sampling, we can estimate the standard error for $\hat{\Gamma}$ and construct a large-sample confidence interval for $\Gamma$. 
The delta method can approximate the standard error. 
$\sqrt[]{\mathstrut n}(\hat{\Gamma} - \Gamma)$ has an asymptotic (as $n \rightarrow \infty$) normal distribution with mean zero and variance $\sigma^2[ \Gamma ]$.
See Appendix 2 for the details of $\sigma^2[ \Gamma ]$.

Let $\hat{\sigma}^2[ \Gamma ]$ denote $\sigma^2[ \Gamma ]$ where \{$p_{ij}$\} is replaced by \{$\hat{p}_{ij}$\}.
Then $\hat{\sigma} [ \Gamma ]/\sqrt[]{\mathstrut n}$ is the estimated approximate standard error for $\hat{\Gamma}$, and 
$\hat{\Gamma} \pm z_{p/2} \hat{\sigma} [ \Gamma ]/\sqrt[]{\mathstrut n}$ is an approximate $100(1-p)$ percent confidence interval for $\Gamma$,
where $z_{p/2}$ is the $100 (1-p/2)$th percentile of the standard normal distribution.

The asymptotic normal distribution may not be applicable when estimating measures on small sample datasets. 
In small dataset, the sample proportion of $(i,  j)$ cell may fall 0 (i.e., $\hat{p}_{ij} = 0$).
Thus, we consider Bayesian methods. 
Although the sample proportion is typically used to estimate the approximate standard error for $\hat{\Gamma}$, herein we consider the Bayes estimator derived from the uninformed prior probability. 
To have a vague prior, the Haldane prior is used for the prior information (see Haldane 1932; Berger 1985, p.89). 
We set all parameters of the Dirichlet distribution to 0.0001 when estimating the approximate variance of the proposed measure.
\\

\noindent \textbf{\large 4. Data analysis}

\noindent \textbf{\large 4.1. Artificial data}

\noindent \textbf{\large 4.1.1. Role of distance postulates for visualization}

To illustrate the concept of visualization, we used artificial datasets in two scenarios: one that satisfies the structure of MH and one that has location-shifted marginal distributions. 

The visualization in Table 4(a) shows that all values of sub-measure $\hat{\gamma}_i$ are equal to zero, and the value of the proposed measure $\hat{\Gamma}$ is zero (i.e., the MH model holds). 
In terms of information divergences, the two marginal distributions can be interpreted as the same. 
Therefore, the values of the label, which is the sub-measure using the Matusita distance, are zero, and points are drawn at $\left( \frac{1}{2}, \frac{1}{2} \right)$ in the visualization (Figure 3(a)). 

The visualization in Table 4(b) shows that all values of sub-measure $\hat{\gamma}_i$ are equal to 0.341 because the assumed structure shows location-shifted marginal distributions. 
Since we estimated $\left(\hat{G}^c_{1(i)} < \hat{G}^c_{2(i)}\right)$, the point on the graph is drawn from $\left( \frac{1}{2}, \frac{1}{2} \right)$ to the upper left (Figure 3(b)). 
Because the label values are sub-measures using the Matusita distance that satisfies distance postulate (ii), it can be interpreted as the distance between $\left( G^c_{1(i)}, G^c_{2(i)} \right)$ and $\left( \frac{1}{2}, \frac{1}{2} \right)$.
However, the direction is crucial when using the Kullback-Leibler divergence (see Appendix 1). 
When using the Kullback-Leibler divergence in Table 4(b), the distance from $\left( \frac{1}{2}, \frac{1}{2} \right)$ to $\left( G^c_{1(i)}, G^c_{2(i)} \right)$ and the distance from $\left( G^c_{1(i)}, G^c_{2(i)} \right)$ to $\left( \frac{1}{2}, \frac{1}{2} \right)$ differ (Table 5). 
Therefore, the label value must be selected carefully because this divergence may hinder an intuitive interpretation. 
In addition, it can be evaluated appropriately in indirect comparisons between two points for the distance from a reference since the proposed measure satisfies the triangular inequalities.

Thus, the visualization must use a divergence that satisfies the distance postulates. 
In addition, the proposed visualization gives a natural and intuitive interpretation because we can understand the degree of departure from MH for each level $i$, and the sub-measure calculated by $\hat{G}^c_{1(i)}$ and $\hat{G}^c_{2(i)}$  compares the marginal cumulative distributions $\left( \hat{F}_{1(i)}~{\rm and}~\hat{F}_{2(i)}  \right)$.
This section shows the visualization in monotonic differences of the marginal cumulative distributions, but the next section illustrates the relationship between marginal cumulative distributions and visualizations in several patterns.
\begin{table}[h]
\small\sf\centering
\caption{Artificial data.\label{T4}}
\begin{tabular}[!bh]{ccccc} \hline
	& 1	& 2	& 3	& 4	\\ \hline
\multicolumn{5}{l}{(a)}\\ 
1	& 0	& 10	& 10	& 10	\\
2	& 10	& 0	& 10	& 10	\\
3	& 10	& 10	& 0	& 10	\\
4	& 10	& 10	& 10	& 0	\\ \hline
\multicolumn{5}{l}{(b)}\\ 
1	& 0	& 10	& 10	& 10	\\
2	& 30	& 0	& 10	& 10	\\
3	& 30	& 30	& 0	& 10	\\
4	& 30	& 30	& 30	& 0	\\ \hline
\end{tabular}
\end{table}
\begin{figure}[h]
\centering
\includegraphics{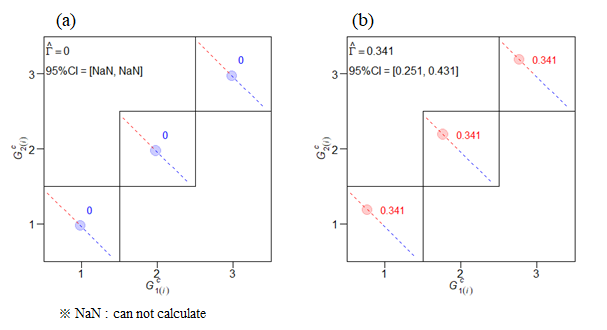}
\caption{Visualization result of Table 4.}
\end{figure}
\begin{table}[h]
\small\sf\centering
\caption{Values of the Kullback-Leibler divergence.\label{T5}}
\begin{tabular}[!bh]{ccccc} \hline
$i$	& $\hat{G}^c_{1(i)}$	& $\hat{G}^c_{2(i)}$	& $\hat{K}^1_i$	& $\hat{K}^2_i$	\\ \hline
1	& 0.250		& 0.750		& 0.131		& 0.144		\\
2	& 0.250		& 0.750		& 0.131		& 0.144		\\
3	& 0.250		& 0.750		& 0.131		& 0.144		\\ \hline
\multicolumn{5}{l}{$\hat{K}^1_i$ = $I^{(0)}_i \left( \left\{ \hat{G}^c_{1(i)}, \hat{G}^c_{2(i)} \right\} ; \left\{ \frac{1}{2}, \frac{1}{2} \right\} \right)$} \\
\multicolumn{5}{l}{$\hat{K}^2_i$ = $I^{(0)}_i \left( \left\{ \frac{1}{2}, \frac{1}{2} \right\} ; \left\{ \hat{G}^c_{1(i)}, \hat{G}^c_{2(i)} \right\} \right)$} \\
\end{tabular}
\end{table}
\\

\noindent \textbf{\large 4.1.2. Perception of different behaviors between categorical levels}

Our visualization can interpret the relationships between the marginal cumulative distributions, which is difficult using a single-summary-measure. 
Here, we treat artificial data where the values of the single-summary-measure are the same, but the visualizations of the sub-measures behave differently. 
Tables 6(a)–(d) show the artificial data, which are setup so that the value of the measure is 0.341. 
Figures 4(a)–(d) show the visualizations of Tables 6(a)–(d).

Table 6(a) illustrates a scenario where the marginal cumulative distribution is location-shifted constantly. 
This structure would be expected based on the value of the measure. 
In a clinical study, assuming such a situation implies a constant treatment effect from pre-treatment to post-treatment.

In contrast, Table 6(b) represents a scenario where the marginal cumulative distribution spreads as the categorical level $i$ increases.
Moreover, Tables 6(c)–(d) show situations where the marginal cumulative distribution differs at the categorical level $i$. 
In a clinical study, assuming such a situation suggests that the treatment effect depends on the pre-intervention condition.
\begin{table}[h]
\small\sf\centering
\caption{Artificial data.\label{T6}}
\begin{tabular}[!bh]{ccccc} \hline
	& 1	& 2	& 3	& 4	\\ \hline
\multicolumn{5}{l}{(a)}\\ 
1	& 0	& 30	& 30	& 30	\\
2	& 10	& 0	& 30	& 30	\\
3	& 10	& 10	& 0	& 30	\\
4	& 10	& 10	& 10	& 0	\\ \hline
\multicolumn{5}{l}{(b)}\\ 
1	& 0	& 5	& 5	& 6	\\
2	& 5	& 0	& 11	& 36	\\
3	& 5	& 10	& 0	& 86	\\
4	& 5	& 10	& 10	& 0	\\ \hline
\multicolumn{5}{l}{(c)}\\ 
1	& 0	& 30	& 30	& 30	\\
2	& 10	& 0	& 0	& 30	\\
3	& 10	& 240	& 0	& 30	\\
4	& 10	& 10	& 10	& 0	\\ \hline
\multicolumn{5}{l}{(d)}\\ 
1	& 0	& 30	& 30	& 30	\\
2	& 10	& 0	& 30	& 30	\\
3	& 10	& 10	& 0	& 0	\\
4	& 10	& 10	& 160	& 0	\\ \hline
\end{tabular}
\end{table}
\begin{figure}[h]
\centering
\includegraphics{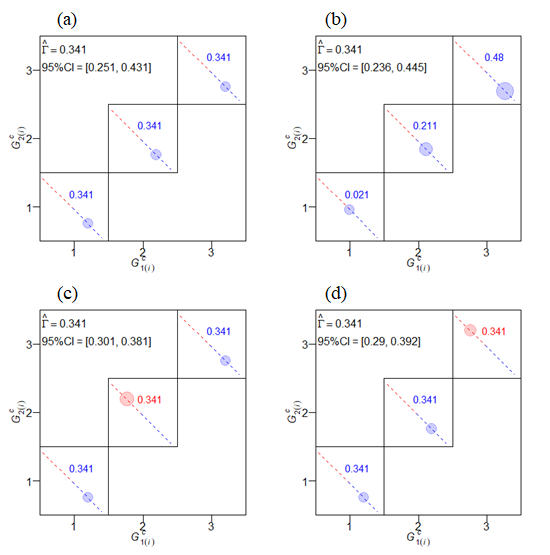}
\caption{Visualization result of Table 6.}
\end{figure}
\\

\newpage
\noindent \textbf{\large 4.2. Simulation studies}

Monte Carlo simulations were performed to theoretically derive the coverage probabilities of the approximate $95\%$ confidence intervals assuming random sampling of an underlying bivariate normal distribution. 
Here, we considered random variables $Z_1$ and $Z_2$ with means ${\rm E}(Z_1) = 0$ and ${\rm E}(Z_2) = d$, variances ${\rm Var}(Z_1) = {\rm Var}(Z_2) = 1$, and correlation ${\rm Corr}(Z_1, Z_2 ) = 0.2$. 
Assuming a $6 \times 6$ table is formed using the cutoff points for each variable at $-1.2, -0.6, 0, 0.6, 1.2$, we evaluated several simulation scenarios where $d = 0.00~{\rm to}~4.00$ by 0.25 and $n = 36, 180, 360, 3600$ (${\rm sparseness~index}=1, 5, 10, 100$). 
The simulation studies were performed based on 100,000 trials per scenario.

Figure 5 plots the mean of random variable $Z_2$ along with the true value of the measure based on a bivariate normal distribution. 
When $d=0$, the true value of the measure is observed as 0 because there is no difference in the means whose condition is stronger than the structure of the MH.
Although the true value increases monotonically for $d=0, \ldots, 1$, a large mean difference between random variables is necessary for the true value to reach 1.
\begin{figure}[h]
\centering
\includegraphics{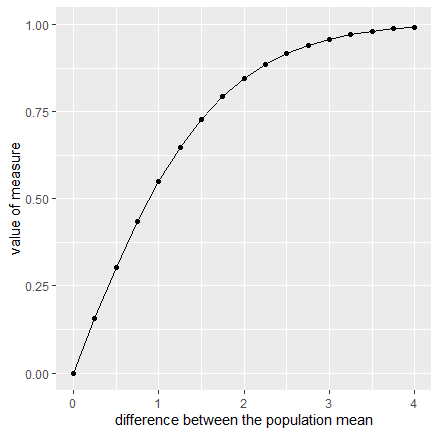}
\caption{Mean of random variable $Z_2$ and the true value of the measure, which are based on a bivariate normal distribution.}
\end{figure}

Figure 6 shows the coverage probability according to the true values. 
For a small sample size, it is difficult to obtain a nominal coverage probability, whereas the coverage probability is maintained at a $95\%$ confidence interval for a sufficient sample size.
\begin{figure}[h]
\centering
\includegraphics{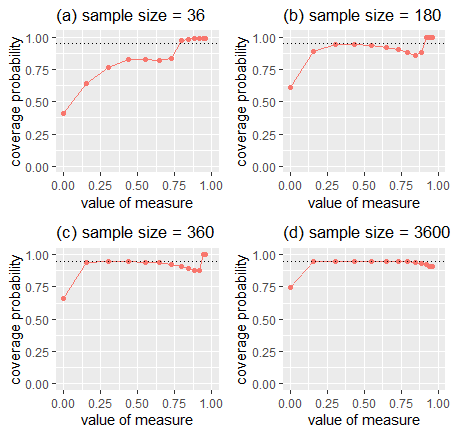}
\caption{Transitions of true values and coverage probabilities.}
\end{figure}
\\

\newpage
\noindent \textbf{\large 4.3. Example}

As an example, consider the data in Table 1. 
In the original work (Sugano $et~al$., 2012), the proportion of improvement or deterioration for the esomeprazole group (drug group) and placebo group were described. 
Table 7 shows the results of applying the proposed measure $\Gamma$ to these data to statistically consider the treatment effects for the drug or placebo. 
The estimate of asymptotic variance using the sample proportion cannot be calculated because $\hat{G}^c_{1(4)}=0$ in Table 1(b). 
Hence, a Bayes estimator is used to estimate the asymptotic variance. 
The $95\%$ confidence intervals do not cross zero, suggesting that both groups have a higher degree of deviation from MH. 
That is, the marginal distribution after the treatment shifts compared to that before the treatment.

For an intuitive understanding, Figure 7 plots the trend, where blue indicates an improving trend and red a deteriorating one. 
The drug group shows an improving trend ($\hat{G}^c_{1(i)} \geq \hat{G}^c_{2(i)}$), while the placebo group displays a deteriorating trend ($\hat{G}^c_{1(i)} < \hat{G}^c_{2(i)}$). 
For the drug group, $i=1, 2, 3$ show an improvement trend, while $i=4$ shows a deteriorating trend although the circle is small (i.e., the proportion of observed frequencies comprising $\hat{G}^c_{1(4)}$ and $\hat{G}^c_{2(4)}$ is small relative to the total). 
These results imply that there might be differences in treatment effects between $i$ levels.
\begin{table}[h]
\small\sf\centering
\caption{Estimates of the measure, approximate standard error, and approximate $95\%$ confidence intervals applied to the data in Table 1.\label{T7}}
\begin{tabular}[!bh]{cccc} \hline
Applied	& Estimated	&Standard	& Confidence	\\
data		& measure	&error		&  interval		\\ \hline
Table 1(a)	& $0.308$	& $0.078$	& $(0.156, 0.460)$	\\
Table 1(b)	& $0.511$	& $0.059$	& $(0.395, 0.627)$	\\ \hline
\end{tabular}
\end{table}
\begin{figure}[h]
\centering
\includegraphics{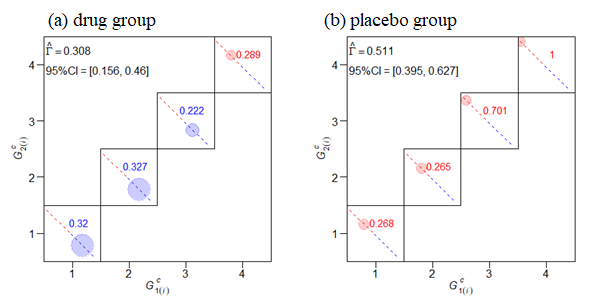}
\caption{Visualization results of Table 1.}
\end{figure}
\\

\noindent \textbf{\large 5. Discussion}

In the proposed measure, sub-measures are used in the visualization to capture features overlooked by a single summary measure. 
Previous studies have adopted similar approaches, except that the sub-measures are not used for interpretation (Tomizawa $et~al.$, 2003; Yamamoto $et~al.$, 2011). 
This study demonstrates that sub-measures allow two kinds of marginal inhomogeneities to be visualized, providing a more detailed interpretation of the single-summary-measure.

The proposed visualization is analyzed using Table 1. 
First, because the Matusita distance satisfies the distance postulates, the visualization that draws points on two-dimensional coordinates can give a natural and intuitive interpretation. 
In particular, the values of existing measures based on the power-divergence (Kullback-Leibler divergence) that do not satisfy distance postulate (ii) would give different values if the distance from the start point to the end point is swapped. 
That is, the data in Table 1 would create two visualization patterns. 
In contrast, for the Matusita distance, the same value is obtained even if the distance from the start point to the end point is swapped. 
Hence, a special annotation is unnecessary for a visual interpretation. 

Furthermore, the point in Figure 7(a) where $i=1, 2, 3$ and $i=4$ show different directions is difficult to discern using the existing measure proposed by Yamamoto $et~al.$ (2011) because it is a single-summary-measure. 
However, the different directions can be considered intuitively through visualization by level $i$.
The proposed visualization does not draw the points on one coordinate because the degree of departure from MH is likely the same for each level in real data analysis (Figure 7). 
This is because identifying which level $i$ of points is drawn is difficult. 
Hence, it is important to satisfy the distance postulates and to consider methods for visualizing categorical data of square contingency tables.

The visualization program was implemented in the R programming language (R Core Team, 2023). 
Noting that a graphical layout in package ``ggplot2'' is defined by ``gtable'' (and also ``grid''). 
In addition, the arrangement of multiple figure objects can be set by package ``gridExtra''.
We used ``grid'' and ``gridExtra'' packages for visualization purposes. 
We referenced the function ``agreementplot()'' by the ``vcd'' package, which is the categorical data visualization package for the ``observer agreement chart''.
\\

\noindent \textbf{\large 6. Conclusion}

The proposed measure $\Gamma$ is the weighted sum of the sub-measures that satisfy all three distance postulates. 
Here, we demonstrate the approximated confidence interval for $\Gamma$.
The proposed visualization using the Matusita distance provides a natural visual interpretation of MH in a square contingency table. 
In addition, we show that the visualization can provide useful interpretations using an example.
\\

\newpage
\noindent \textbf{\large Appendix 1}

Assuming that $\{ G_{1(i)} + G_{2(i)} \neq 0 \}$, the power-divergence-type measure representing the degree of departure from MH proposed by Tomizawa, Miyamoto and Ashihara (2003) for $\lambda > -1$ is given as
\begin{equation}
\begin{split}
\Phi^{(\lambda)} &= \frac{\lambda(\lambda+1)}{2^{\lambda}-1} \sum^{r-1}_{i=1} \left( G^{\ast}_{1(i)} + G^{\ast}_{2(i)}\right) \\
& ~~~~~~~~~~~~~~~ \times I^{(\lambda)}_i \left( \left\{ G^c_{1(i)}, G^c_{2(i)} \right\} ; \left\{ \frac{1}{2}, \frac{1}{2} \right\} \right), \nonumber
\end{split}
\end{equation}
where
\begin{equation}
\begin{split}
I^{(\lambda)}_i (\cdot, \cdot) &= \frac{1}{\lambda(\lambda+1)} \Biggl[ G^c_{1(i)} \left\{ \left( \frac{G^c_{1(i)}}{1/2}\right)^{\lambda} - 1 \right\} \\
& ~~~~~~~~~~~~~~~~~~~~~ + G^c_{2(i)} \left\{ \left( \frac{G^c_{2(i)}}{1/2}\right)^{\lambda} - 1 \right\} \Biggr], \nonumber
\end{split}
\end{equation}
and the value at $\lambda=0$ is taken to the limit as $\lambda \rightarrow 0$. 
Note that $I^{(\lambda)}_i (\cdot, \cdot)$ is the power-divergence between two distributions (see Cressie and Read, 1984; Read and Cressie, 1988, p.15).
Namely, 
\[
I^{(0)}_i (\cdot, \cdot) = G^c_{1(i)} \log \left( \frac{G^c_{1(i)}}{1/2} \right) + G^c_{2(i)} \log \left( \frac{G^c_{2(i)}}{1/2} \right).
\]
This measure has the following characteristics: 
\begin{enumerate}
\item[(i)] $\Phi^{(\lambda)} = 0$ if and only if the MH model holds
\item[(ii)] $\Phi^{(\lambda)} = 1$ if and only if the degree of departure from MH is a maximum, in the sense that  $G^c_{1(i)}=1$ (then $G^c_{2(i)}=0$) or $G^c_{1(i)}=0$ (then $G^c_{2(i)}=1$), for $i = 1, \ldots, r-1$
\end{enumerate}

Second, assuming that $\{ G_{1(i)} + G_{2(i)} \neq 0 \}$, the measure representing two kinds of marginal inhomogeneities proposed by Yamamoto, Ando and Tomizawa (2011) is given as
\[
\Psi = \frac{4}{\pi} \sum^{r-1}_{i=1} \left( G^{\ast}_{1(i)} + G^{\ast}_{2(i)}\right) \left( \theta_i - \frac{\pi}{4} \right),
\]
where
\[
\theta_i = \cos^{-1} \left( \frac{G_{1(i)}}{\sqrt{ G^2_{1(i)} + G^2_{2(i)} }} \right).
\]
This measure has the following characteristics: 
\begin{enumerate}
\item[(i)] $\Psi = -1$ if and only if there is a structure of maximum upper-marginal inhomogeneity
\item[(ii)] $\Psi = 1$ if and only if there is a structure of maximum lower-marginal inhomogeneity
\item[(iii)] If the MH model holds then $\Psi = 0$, but the converse does not hold
\end{enumerate}
Yamamoto $et~al.$ (2011) defined this structure ($\Psi = 0$) as the average MH model.

Third, assuming that $\{ G_{1(i)} + G_{2(i)} \neq 0 \}$, the two-dimensional measure that can simultaneously analyze the degree and directionality of departure from MH proposed by Ando, Noguchi, Ishii and Tomizawa (2021) is given as
\[
\boldsymbol{\tau} = \begin{pmatrix} \Phi^{(0)} \\ \Psi \end{pmatrix}.
\]
This two-dimensional measure has the following characteristics: 
\begin{enumerate}
\item[(i)] $\boldsymbol{\tau} = (0, 0)^t$ if and only if the MH model holds
\item[(ii)] $\boldsymbol{\tau} = (1, -1)^t$ if and only if there is a structure of maximum upper-marginal inhomogeneity
\item[(iii)] $\boldsymbol{\tau} = (1, 1)^t$ if and only if there is a structure of maximum lower-marginal inhomogeneity
\end{enumerate}

\newpage
\noindent \textbf{\large Appendix 2}

Using the delta method, $~\sqrt[]{\mathstrut n}(\hat{\Gamma}  - \Gamma)$ has an asymptotic variance $\sigma^2[ \Gamma ]$, which is given as 
\[
\sigma^2[ \Gamma ]  = \sum^{r-1}_{k=1} \sum^{r}_{l=k+1}  \left( p_{kl} D^2_{kl} + p_{lk} D^2_{lk} \right),
\]
where
\begin{equation}
\begin{split}
D_{kl} &= \frac{1}{\Delta} \sqrt{ \frac{2+\sqrt{2}}{2} } \sum^{r-1}_{i=1} I(k \leq i, l \geq i+1) A_i - \frac{(l-k)}{\Delta} \Gamma, \\
D_{lk} &= \frac{1}{\Delta} \sqrt{ \frac{2+\sqrt{2}}{2} } \sum^{r-1}_{i=1} I(k \leq i, l \geq i+1) B_i - \frac{(l-k)}{\Delta} \Gamma, \\
A_i &= \frac{1}{2\sqrt{C_i}} \left(  2C_i + \upsilon_{1(i)} \frac{G^c_{2(i)}}{\sqrt{G^c_{1(i)}}} - \upsilon_{2(i)} \sqrt{G^c_{2(i)}} \right), \\
B_i &= \frac{1}{2\sqrt{C_i}} \left(  2C_i - \upsilon_{1(i)} \sqrt{G^c_{1(i)}} + \upsilon_{2(i)} \frac{G^c_{1(i)}}{\sqrt{G^c_{2(i)}}} \right), \\
C_i &= \upsilon_{1(i)}^2 + \upsilon_{2(i)}^2, \nonumber
\end{split}
\end{equation}
and $I(\cdot)$ is indicator function.

\end{document}